\RequirePackage{fix-cm}
\documentclass[smallextended]{svjour3}       
\smartqed  
\usepackage{amsmath}
\usepackage{graphicx}
\usepackage{xcolor}
\usepackage{comment}
\begin{document}

\title{Visual Representation for Patterned Proliferation of Social Media Addiction: Quantitative Model and Network Analysis} 

\subtitle{}

\titlerunning{A Quantitative Model of Patterned Proliferation on Social Media Addiction}        

\author{Dibyajyoti Mallick, Priya Chakraborty, Sayantari Ghosh 
}


\institute{Dibyajyoti Mallick$^*$ \at
              Department of Physics, National Institute of Technology Durgapur \\
              \email{dm.20ph1103@phd.nitdgp.ac.in}           
           \and
           Priya Chakraborty$^*$ (corresponding author)\at
              Department of Physics, National Institute of Technology Durgapur \\
              \email{pc.20ph1104@phd.nitdgp.ac.in}
              \and
           Sayantari Ghosh \at
              Department of Physics, National Institute of Technology Durgapur \\
              \email{sayantari.ghosh@phy.nitdgp.ac.in}
              \and
              $^*$ DM and PC share equal credit for this research.
}

\date{Received: date / Accepted: date}

\maketitle

\providecommand{\keywords}[1]
{
  \small	
  \textbf{\textit{Keywords---  Epidemic models, Social media addiction, Complex networks, Bifurcation analysis, Pattern formation}}
}

\begin{abstract}
With the advancement of information technology, more people, especially young adults, are getting addicted to the use of different social media platforms. Despite immense useful applications in communication and interactions, the habit of spending excessive time on these social media platforms is becoming addictive, causing different consequences, like anxiety, depression, health problems, and many more. Here, we mathematically explored a model of social media addiction, including a peer-influence relapse. We have further done the complex network analysis for a heterogenic synthetic society. Finally, we explore spatiotemporal pattern formation in a diffusive social system using the reaction-diffusion approach. Our model shows how the existent nonlinearity in the system makes it difficult to make society social media addiction free once it crosses a certain threshold. Some possible strategies are explored mathematically to prevent social media addiction, and the importance of peer-influenced relapse has been identified as a major barrier. 
\end{abstract} \hspace{10pt}

\section{Introduction}

As the era of information technology and the Internet has emerged, the current decade has seen exponential growth in the number of individuals with online social media presence (in platforms like Facebook, YouTube, WeChat, Instagram, etc.). The possibilities of personal interactions and communications have seen a drastic proliferation, causing a huge change in the amount of time an average adult spends online \cite{smith2018social,clement2020daily}. Recent studies show that in July 2020, 4 billion adults were active as social media users worldwide, with the number consistently growing every day \cite{verduyn2020social,meral2021social}. More than 73\% and 68\%  of the adult population of the U.S. \cite{smith2018social} are attracted to Facebook and YouTube respectively. The immense power of social media to create and maintain connections beyond geographical or time constraints comes with the consequence of Social media addiction (SMA). This can be regarded as one form of urge and compulsion to use one or multiple social media platforms regularly and frequently, without which the individual suffers from psychosocial distress \cite{griffiths2000does,starcevic2013internet}. \\
The indications of SMA echoes in mood swings, perception problems, reduced bodily reactions, and face-to-face communication problems \cite{balakrishnan2013malaysian,blachnio2019cultural,blachnio2017role,kuss2011online,verduyn2021impact,omar2020influence}. The direct consequences of SMA are associated with several issues arising related to emotional torment, relational turmoil, health problems, and performance  (e.g.,\cite{echeburua2010severe,kuss2011online,marino2017objective,marino2018associations}). Researchers have found, for example, that in Central Serbia \cite{pantic2012association} and in the United States, time invested in social media correlates positively to symptoms of depression among young adults. Excess and compulsive use of social media has been linked with a diminished performance at job \cite{j2014internet,xanidis2016association,saputri2023social}, lowered academic grades \cite{al2015dimensions}, unhealthy social connections \cite{fox2015dark}, sleep disorders \cite{koc2013facebook,wolniczak2013association}, depression, and dissatisfaction \cite{przepiorka2016time}, and negative feelings like mistrust, nervousness, and despair \cite{elphinston2011time,pantic2014online}. \\
Over the past years, gradually, the global public concern related to SMA has also increased, and interesting underlying factors like the role of peer influence in SMA have started being explored. However, these social media platforms are commonly considered an effective avenue for communication among peers, only recently \cite{wegmann2016internet,saqib2022social} a peer-influence perspective on compulsive social networking site users has been explored with a closer look. Internet-based contagions like false information spreading, viral marketing, viral effects meme, etc., are common nowadays, and these have also been explored in the light of epidemic models \cite{zhao2012sihr,bhattacharya2019viral,ghosh2020ensuring,wang2011epidemiological}. Moreover, the role of peer influence in the development of other harmful, compulsive, and addictive behaviors (e.g., substance use, alcohol addiction, smoking, gaming addiction, etc.) \cite{liu2022social} has been observed. As many such addictive behaviors have also been studied with the help of mathematical models \cite{huang2013note,fang2015global,sanchez2007drinking,lahrouz2011deterministic,gaurav2022purchase}, building up a quantitative understanding of the major causes and possible remedies of social media addiction using tools of mathematical epidemiology is thus of utmost importance.\\ 
It has been established following the social identity theory, people generally feel obliged to raise his/her online activity by conforming to group norms, following their peer/group’s footsteps. Individuals are expected to react/respond to their friend’s posts, which gradually pushes them to engage in obligatory social media usage \cite{turel2019social}. This is the negative peer influence, which has been included in a recently explored mathematical model in this context \cite{alemneh2021mathematical}. However, the role of this negative peer influence cannot be ignored in the relapse process as well. In another recent work, Turel et al., have explored the phenomena of short abstinence of social media use and relapse \cite{turel2019social}.  They have drawn similarities with other addictions like smoking and drinking, establishing that the harmful push from their online peers may play a significant role in the relapse of SMA. Keeping this in mind, in this work we propose a mathematical model of this dynamics to identify major drivers and barriers of this social dynamics, including a peer-influenced relapse. We also explore the model beyond the deterministic scenario, to consider the heterogeneity of society, and determine the dynamics of a synthetic society using complex networks. Moreover, considering a reaction–diffusion counterpart of the proposed SMA model, we explore the dynamics with visual understanding. With some recent studies of Turing Pattern analysis for rumor dynamics systems \cite{zhu2022pattern}, we explore the transient dynamics for SMA considering the spatial distribution of addictive behavior\cite{spertus2005evaluating}. The rest of the paper is organized as this: in Section $2$, we have elaborated on model formulation; in Section $3$, we have done the bifurcation analysis; in Section $4$ we have further analyzed the model on a complex network while in Section $5$ we have explored pattern formation considering the diffusion in a two-dimensional space for randomized population. Finally, we conclude with a brief discussion in Section $6$.

\section{Model Formulation}
\begin{figure}
    \centering
    \includegraphics[width=0.6\linewidth]{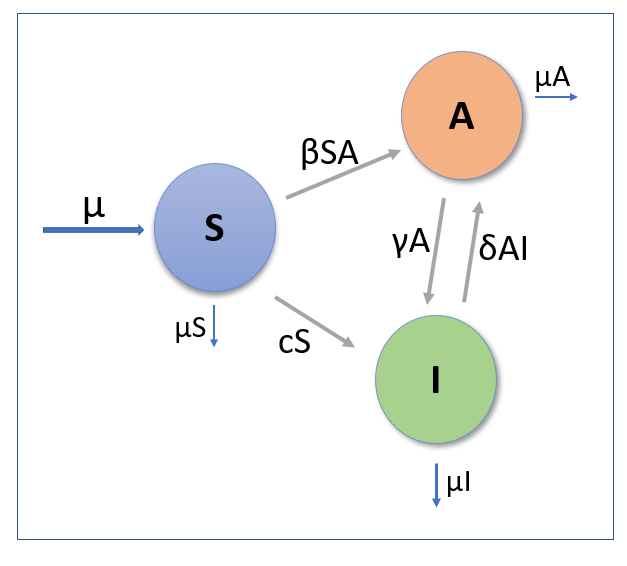}
    \caption{Block diagram of the proposed model of social media addiction. The notation for compartments (denoted by solid circles) and transition rates (denoted by arrows with parameters) have been elaborated in the text. }
    \label{scheme}
\end{figure}

To understand the SMA dynamics quantitatively, we have taken a compartmental epidemic model approach. A set of ordinary differential equations has been used to define the possible transitions between different social compartments. The proposed compartmental model is shown in Fig. \ref{scheme}. At any time $t$, the closed total population $N(t)$ is subdivided into three compartmental states: Susceptible($S$), Addicted($A$), Inert($I$).
\begin{itemize}
    \item Susceptible($S$) $:$ As the name suggests, in this group, people are not addicted yet to the use of any social media applications, or websites. They may use social media less frequently but have not grown any compulsive behavior. Over time they may get influenced by social media-addicted peers and could spend an unjustified amount of time online. Each person in this population is denoted as $S$. 
    
   \item Addicted($A$) $:$ People in this group are addicted to social media and spend most of their time on it. Each person in this class is represented by $A$.     
  
   \item Inert($I$) $:$ People in this group got over their SMA and have grown some ability to control their online actions. People in this class are also influencing other people to join them. However, some of them may not be rigid about their decision, and they can again go back to the addicted state by getting influenced by their addicted peers. Each person in this class is denoted by $I$.
\end{itemize} 
For this analysis, we consider the total population to be normalized to $1$, in which a demographic variation has been allowed. In this population, $\mu$ is the rate by which people enter the dynamical system. At the same time, in the normal course, some people can leave that population from each state with the effective term, $\mu$$S$, $\mu$$A$, or $\mu$$I$, keeping the population size intact. Thus at any point in time, we have:

\begin{equation*}
     S(t)+A(t)+I(t) = 1
 \end{equation*}
 Now, let us discuss all the possible transitions one by one considered in the model, categorically. Following possibilities of transitions may exist as each individual from one subpopulation can change their state to take another state, over a given time.
\begin{itemize}
    \item {\textbf{Possible transitions in class $S$ :}} The effective contact rate by which susceptible individuals are getting addicted to social media due to the influence of addicted individuals is $\beta$; thus, per unit time the amount $\beta$$SA$ will get subtracted from this class and should be added to the rate equation of $A$. Some people from $S$ who understand the consequences of heavy social media presence, may quit or substantially reduce their usage, and directly join the Inert class at a rate $c$. Thus, the term $cS$ will be subtracted from $S$ and added to the rate equation of $I$. 
   \item {\textbf{Possible transitions in class $A$ :}} As mentioned in the previous case, $\beta$$SA$ must get added to this class due to the increase in the number of social media addicts. Here $\gamma$ is the rate by which addicted people are becoming inert by quitting the use of social media. Development of self-awareness among people regarding SMA from various local as well as global factors may be responsible for this. So, the $\gamma$$A$ term should be deducted from this class and get added to $I$. At the same time, those who are not rigid about their decision can again join this class with rate $\delta$ so $\delta$$AI$ amount must be added here and subtracted from $I$. Here the negative peer influence on relapse has been considered. 
   \item {\textbf{Possible transitions in class $I$ :}} As discussed in possible changes in class $S$ and $A$, $\gamma$$A$, $cS$ should be added in this class, while $\delta$$AI$ will be deducted from the rate equation of this class.  
\end{itemize}
Thus the resulting equations describing these dynamics are given by:
\begin{eqnarray}
\frac{dS}{dt}=\mu - \beta\;S\;A - \mu\;S - c\;S \nonumber   \\ 
 \frac{dA}{dt}= \beta\;S\;A-\gamma\;A-\mu\;A+\delta\;A\;I  \label{rkn} \\ 
\frac{dI}{dt}= \gamma\;A-\mu\;I-\delta\;A\;I+c\;S \nonumber
\end{eqnarray}
All parameters should be real and positive. Now, we will analyze the system in a steady state, considering all rates of changes reaching a dynamic equilibrium.
\begin{figure}
    \centering
  \includegraphics[width=0.6\textwidth]{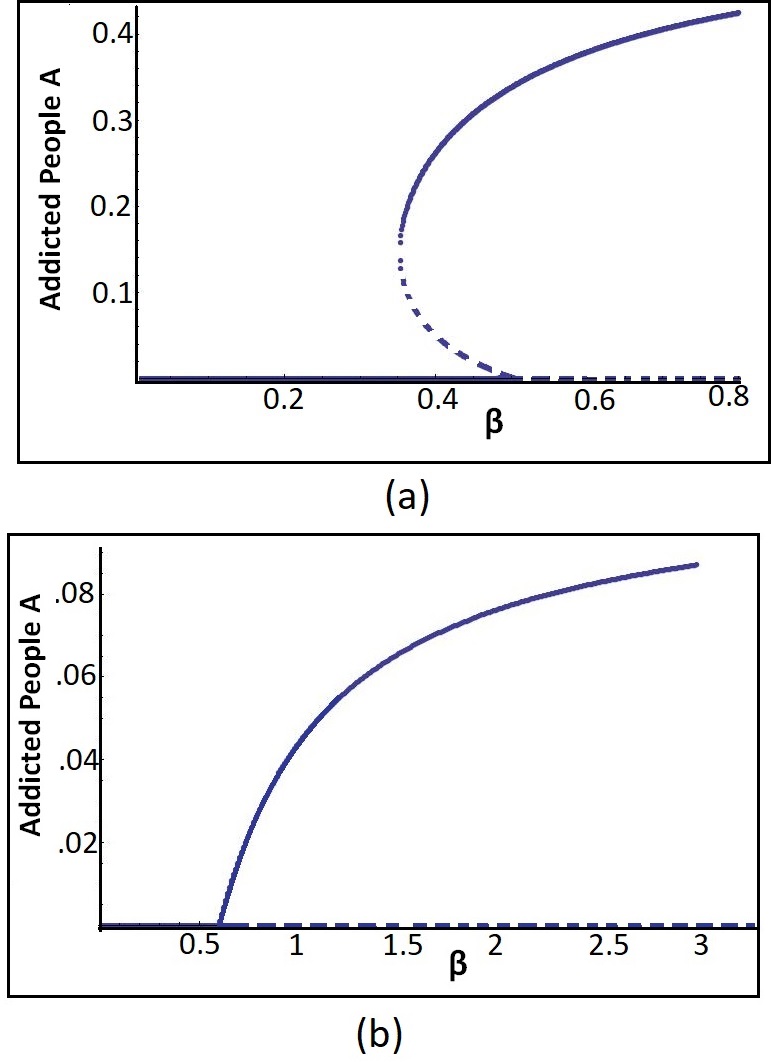}
    \caption{Variation in steady-state population of Addictive people $A$, with respect to the parameter $\beta$. (a) For $\delta=1$, there exists a bistable behavior for a range of $\beta$ values. (b) For $\delta=0.1$ there is only one steady-state solution. An addiction-free equilibrium up to a range of $\beta$ switches to a state of equilibrium of a certain fraction of the addictive population by a forward bifurcation. Stable states are represented with solid lines, unstable states are represented with dashed lines. The rest of the parameter values for both curves are $\mu=0.05,\;\gamma=0.5,\;c=0.005.$ }
    \label{bifurcation}
\end{figure}
\section{Result: Bifurcation Analysis}
To proceed with the steady-state analysis, we equate all the rates of changes of Eqn. \ref{rkn} to \textit{zero}. The first equation of Eqn. \ref{rkn}, the equation for counting the rate of changes in Susceptible people  gives the steady state value of the population (say $S^*$) in terms of parameters of the model and Addicted people $(A^*)$ as 
\begin{equation}\label{steadystates}
    S^*=\frac{\mu}{\beta\;A^*+\mu+c}
\end{equation}
Now, replacing the value of $S^*$ and putting $I^*=(1-S^*-A^*)$ in the system equation we receive a simple quadratic equation:
\begin{equation}
    X\;(A^*)^2+Y\;A^*+Z=0 
\end{equation}
Where:
\begin{equation*}
X=\delta\;\beta,\;\;\;\;\;Y=\gamma\;\beta+\beta\;\mu+\delta\;c+\delta\;\mu-\delta\;\beta,\end{equation*}
\begin{equation*}
Z=\gamma\;c+\gamma\;\mu+\mu^2-\delta\;c-\beta\;\mu.
\end{equation*}
A careful observation shows that $X$ is always positive, and $Y$ is positive for low values of $\delta$. However, the values of $\beta$ and $c$ determine majorly if $Z$ is positive or negative.
\begin{itemize}
    \item For the negative value of $Z$, the system has only one real positive solution, the other one is negative and thus not accepted physically.
    \item For positive value of $Z$, depending upon the sign of $Y$, thus upon the value of the non-linear back transition rate $\delta$, the system poses multiple (for high value of $\delta$) or no solutions (for low value of $\delta$).
\end{itemize}
Fig. \ref{bifurcation} shows the detailed results. We fixed the value of $\mu, \gamma, c $ $(\mu=0.05,\;\gamma=0.5,\;c=0.005)$ and plot the Addicted population fraction with respect to parameter $\beta$ for two different values of $\delta.$ For $\delta=1$, (Fig. \ref{bifurcation}(a)) we can see for a range of $\beta$ value, the system poses \textit {three} solutions, \textit{two} of which are physically achievable as the third one is an \textit{unstable} fixed point. This referred to the bistable nature of the system, where both the social media addicted or inert population can dominate the population, depending upon the initial conditions. The bistable dynamics is very well known for introducing memory in the system, a history-dependent behavior, where in the region of bistability the system recalls its previous state, and refuses to transit to the other steady state for the smaller fluctuations in the parameter value. It can be said that, for the backward bifurcation seen in this dynamics, while reducing the value of $\beta$, the system chooses to stay in its high $A$ value, for the inherent non-linearity present in the dynamics, and it is difficult to remove the addiction from the system, though in time of forward transition the system was in its low addiction (low $A$) state for the same value of $\beta$. While this bistability is valid for a range of $\beta$ value, for larger $\beta$ values, there is only one steady solution for the addicted population, and for lower $\beta$, the steady solution is addiction free.  
\\For $\delta=0.1$, we find in Fig. \ref{bifurcation}(b), there exists only one stable solution in the system. A forward bifurcation determines the transition from an addiction-free steady-state population to a social media-addicted population with an increase in value of $\beta$. We proceed with the analysis of the model in heterogeneous systems to observe the consequences of this bistable response.
\begin{figure}
    \centering
    \includegraphics[width=0.6\textwidth]{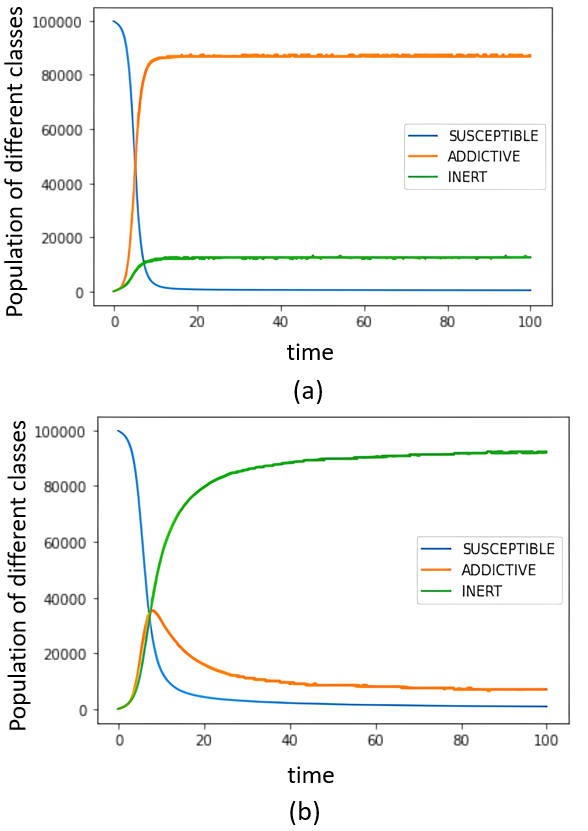}
    \caption{ Variation of different populations concerning time. (a) For $\delta=1$ (b) For $\delta=0.1$. The rest of the parameter values for both curves are $\mu=0.05,\;\gamma=0.5,\;c=0.005$, $\beta=0.36$.}
    \label{network}
\end{figure}

\section{Result: Model analysis on complex network}

In this section, we report the numerical simulation result of our respective SMA model over the network. One of the major issues in ODE-based models is homogeneous mixing, which considers that every people in a population has an equal probability of having contact with every other individual in the considered society. As our society is highly heterogeneous, and to accommodate that fact into our findings, we study the model in the heterogeneous setting of a complex network.\\ 
The simulation is performed on a random network having 100000 nodes with an average degree of 5 and the results are shown in Fig. \ref{network}. Here we have used the EoN module in Networkx on Python to run the simulation in Google Colaboratory. We know Peer influence happens when people are influenced or motivated toward something by watching their friends or relatives. Negative peer influence always sways people toward risky activity and that is observed here in terms of the steady existence of the addictive population. From the deterministic analysis described in the previous section, the relapse parameter $\delta$ is responsible for this negative influence. We can see from Fig. \ref{network}(a) for $\delta$= 1, more people are getting influenced by their friends, and neighbors to use different social media platforms, and the addictive population increases and then eventually saturates after some time, and for $\delta$= 0.1, it increases and attains a peak and then saturates to a much lower value as shown in Fig. \ref{network}(b). It occurs due to the presence of bistability in the system. Meanwhile, the other populations are behaving naturally. An interesting observation here is that though in deterministic analysis for $\delta=0.1$, the steady state value associated with $A$ population was \textit{zero}, here, due to heterogeneity of the synthetic society, in a long time limit, some addicted individuals always persist. Here the final observed fraction is \{$S^*,A^*,I^*$\}= \{0.99706, 0.00233, 0.00061\} instead of \{1,0,0\}, as found from the deterministic analysis. This explains that our society, as a network here, is highly heterogeneous and there are always some people who can become negative peer influencers and some addicted people can always stay. While in a homogeneous setting, no addicted individuals will exist.

\begin{figure}
    \centering
    \includegraphics[width=\textwidth]{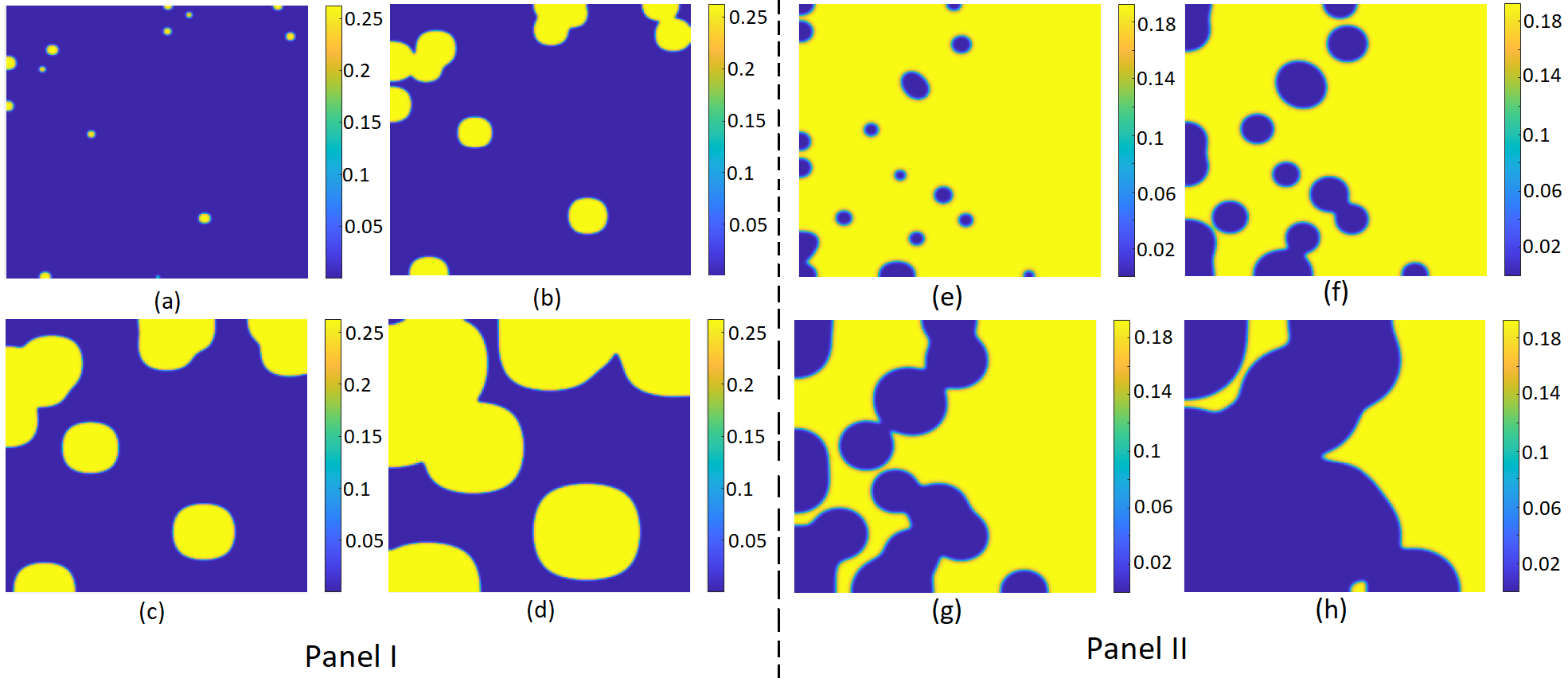}
    \caption{Pattern formation during proliferation of addicted population. Panel I: Parameter values are $\beta=0.4,\;k_1=1,\;k_2=0.1,\;k_3=0.4$.  Diffusion coefficients are taken as $D_S=D_A=D_I=0.01$. Snapshots are taken after the time (a) 600 (b) 2000 (c) 3010 (d) 5000. Panel II: Parameter values are $\beta=0.36,\;k_1=1,\;k_2=0.6,\;k_3=0.4$. Diffusion coefficients are taken as $D_S=D_A=D_I=0.02$. Snapshots are taken after the time (a) 900 (b) 2000 (c) 3000 (d) 5000. Other parameters for both the figures are $\gamma=0.5,\;\delta=1,\;\mu=0.05,\;c=0.005$.}
    \label{diffusion}
\end{figure}

\section{Result: Reaction-diffusion system \& Pattern formation for SMA Dynamics}
\begin{figure}
    \centering
    \includegraphics[width=\textwidth]{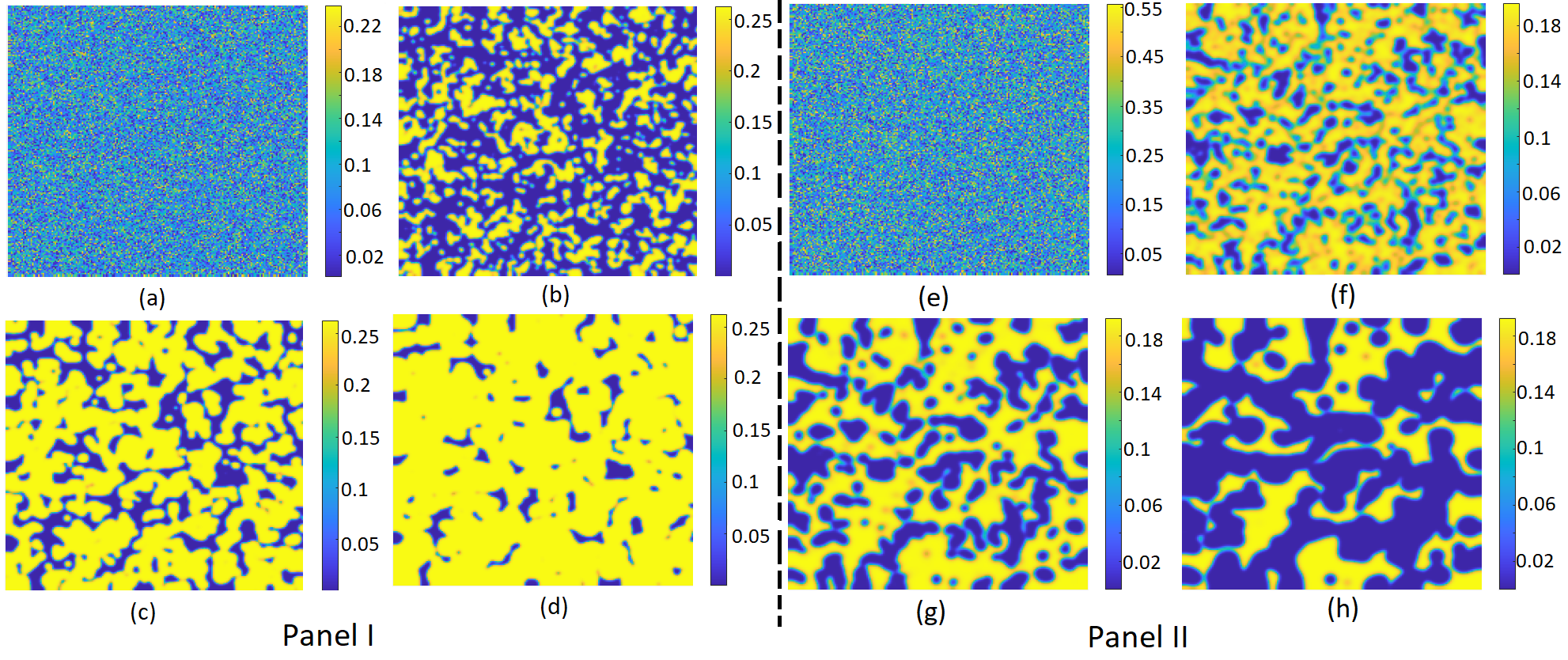}
    \caption{Pattern formation during proliferation of addictive population for different scaling factor value. Panel I: The system, starting from a randomized population, via making a pattern of low and high addiction populations, eventually converges into an addictive population. Parameter values are $\beta=0.4,\;k_1=1,\;k_2=0.2,\;k_3=0.4$.  Diffusion coefficients are taken as $D_S=D_A=D_I=0.01$. Snapshots are taken after the time (a) 1 (b) 200 (c) 410 (d) 600. Panel II: The system, starting from a randomized population, via making a pattern of low and high-addiction populations, eventually converges into an addiction-free population.  Parameter values are $\beta=0.36,\;k_1=1,\;k_2=0.5,\;k_3=0.4$. Diffusion coefficients are taken as $D_S=D_A=D_I=0.02$. Snapshots are taken after the time (e) 1  (f) 160 (g) 315 (h) 630. Rest of the parameter values for both the panel figures are $\gamma=0.5,\delta=1,\;\mu=0.05,\;c=0.005$. }
    \label{pattern}
\end{figure}

To explore the temporal and spatial distribution of the spreading process with visual understanding, we consider that the SMA diffuse in the surrounding of an addicted individual on a 2D lattice population. We consider a $200\times 200$ array of population and allow diffusion in this two-dimensional space for a no-flux boundary condition. In the presence of diffusion Eqn. \ref{rkn} will convert to: 
\begin{equation}
 \frac{\partial S(x,y,t)}{\partial t}=\mu - \beta\;S\;A - \mu\;S - c\;S+D_S^x\;\frac{\partial^2S}{\partial x^2}+D_S^y\;\frac{\partial^2S}{\partial y^2} \nonumber   
\end{equation}
\begin{equation}\label{reacion diffusion}
 \frac{\partial A(x,y,t)}{\partial t}= \beta\;S\;A-\gamma\;A-\mu\;A+\delta\;A\;I + D_A^x\;\frac{\partial^2A}{\partial x^2}+D_A^y\;\frac{\partial^2S}{\partial y^2}   
\end{equation}
\begin{equation*}
    \frac{\partial I(x,y,t)}{\partial t}= \gamma\;A-\mu\;I-\delta\;A\;I+c\;S+D_I^x\;\frac{\partial^2I}{\partial x^2}+D_I^y\;\frac{\partial^2I}{\partial y^2} 
\end{equation*}
The rate of diffusion for population $S$ is taken as $D_S^x$ in $x$ direction and $D_S^y$ along $y$ direction; for population $A$, this rate is taken as $D_A^x$ along $x$ and $D_A^y$ along $y$, and similarly, for population, $I$ the rates of diffusion are taken as $D_I^x$ along $x$ and $D_I^y$along $y$ direction. We have considered further $D_S^x=D_S^y=D_S$, $D_A^x=D_A^y=D_A$, and $D_I^x=D_I^y=D_I$.  \\
Now, let us consider a randomly distributed population, having a distribution profile of all \textit{three} (Susceptible $S$, Addicted $A$, and Inert $I$) populations as given below. 
\begin{eqnarray}
S_{initial}(x,y,0)=k_1\;\zeta(0,1) \nonumber\\
A_{initial}(x,y,0)=k_2\;\zeta(0,1)\\
I_{initial}(x,y,0)=k_3\;\zeta(0,1) \nonumber
\end{eqnarray}
Here, the indices $(x,y,0)$ imply the $x$ and $y$ coordinate position on the lattice and time, respectively. Moreover, the uniform random distribution term $\zeta(0,1)$, along with the respective scaling factors $k_1$, $k_2$, and $k_3$, have been used to randomize the initial population. For different values of these scaling factors and $\beta$, respective pattern formations are shown in Fig. \ref{diffusion}, Fig. \ref{pattern}. \\
In Fig. \ref{diffusion} Panel I and Panel II, it is clearly shown how the parametric dependency of the saddle-node bifurcation determines the fate of the system. Both dynamics are representative for parameter values $\mu=0.05,\;c=0.005,\;\gamma=0.5,\;\delta=1$, but the value of $\beta=0.4$ for Panel I and $\beta =0.36$ for Panel II. We can see that even starting from a very low addictive population Panel I(a), the diffusion eventually spreads the addiction in the total population, and with time eventually, the other populations are also converting to an addictive population (Fig. \ref{diffusion} Panel I(a)-(d)). Conversely, when $\beta$ is low (Fig. \ref{diffusion} Panel II (e)-(h)), the system eventually converges to its low addiction steady state, even starting from a very high population of addicted people.  A similar study with the same $\beta $ and all other parameters as Fig. \ref{diffusion} Panel I (Fig. \ref{diffusion} Panel II) for different scaling factor values are shown in Fig. \ref{pattern} Panel I (Fig. \ref{pattern} Panel II). The similar qualitative output establishes that the steadiness or the fate of the system is entirely dependent on $\beta$ and not upon the scaling factors at all.

\begin{figure}
    \centering
    \includegraphics[width=\textwidth]{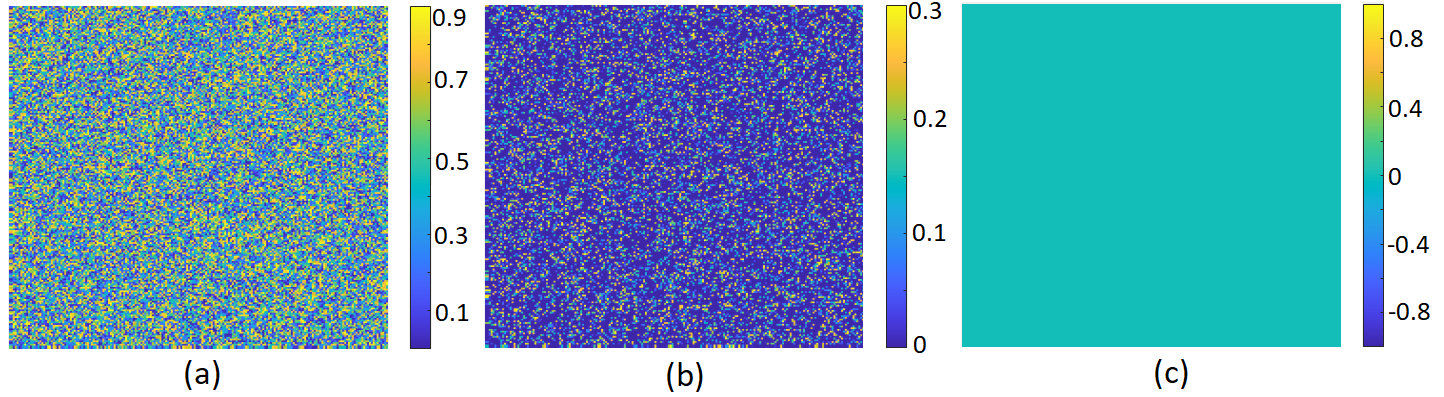}
    \caption{No pattern is formed, and the system quickly converges to an addiction-free equilibrium state. Parameter values are $\gamma=0.5,\delta=0.1,\;\mu=0.05,\;c=0.005,\beta=0.4,\;k_1=1,\;k_2=1,\;k_3=0.4$.  Diffusion coefficients are taken as $D_S=D_A=D_I=0.01$. Snapshots are taken after the time (a) 0.1 (b) 0.8 (c) 1.}
    \label{pitchfork}
\end{figure}

A further study for understanding the transcritical bifurcation is done for the parameter values $\gamma=0.5,\;\delta=0.1,\;\mu=0.05,\;c=0.005,\;\beta=0.4$. For the same parameter values (except $\delta$), the system shows bistability as shown in Fig. \ref{diffusion} Panel I, which eventually converges to its addictive population state for $\delta =1$, while here in Fig. \ref{pitchfork}  for $\delta=0.1$, we get to see that the system quickly converges to its addiction-free equilibrium and no pattern is formed. This reflects the importance of relapse in curbing SMA.

\section{Conclusion}
With the current advent of information technology, social media has gradually gained importance in everyone's life. The slow but steady growth of average time invested by adults and young adults on online social platforms shows our growing dependency on these platforms for communication and entertainment needs. With alarming statistics stating that 210 million individuals are potentially affected by Pathological Internet use, it is becoming increasingly important to prevent this danger, especially for the younger generation. In this paper, we explore the phenomena of social media addiction with a mathematical model to visually understand the implications of addiction spreading with pattern formation on the 2D lattice. For this purpose, we have developed an ODE-based model and analyzed its steady-state behavior to observe the possible presence of saddle-node bifurcation, giving rise to a nonlinear behavior known as bistability. \\
The possible existence of bistability in this dynamics shows the immense sustainability of the addictive state, showing how difficult it could be to get rid of this habit once it has infected a population. We also explore the system with complex networks to compare the similarities and differences in the dynamics of homogeneous and heterogeneous settings. Finally, we study spatiotemporal behavior to show the specific conditions in which the addiction may or may not spread. The spatial behavior of the dynamics shows the emergence of islands and then consequent merging, showing the formation of socially media-addicted smaller groups merging into mega-clusters, which could be prevented by controlling the relapse. We observe that while being addicted in the first place matters, the main problem in curbing SMA is relapse after a break, which should be observed and prevented. As observed in recent work by Soh et al. 2018 \cite{soh2018parents}, the Parent’s intervention in teenagers’ Internet addiction may play a significant role here. They have shown that Parental influence and Parent-child attachment can not only play a significant role in the online activities of their teenagers, but this beneficial parental effect could be stronger than harmful peer influence. This will be explored mathematically in future work to device and validate other possible strategies to prevent social media addiction. Studies of similar nature can also be performed to investigate the dynamics of social contagions, which include not only addictive behaviors, like drug-use behavior, etc. but also social decision-making of individuals depending upon their own ``social bubbles" and peer influence.
\section*{Conflict of Interest}
The authors declare that they do not have any known conflicts of interest.

\newpage
\bibliographystyle{abbrv}
\bibliography{bibliography.bib}

\begin{thebibliography}{10}

\bibitem{al2015dimensions}
J.~J. Al-Menayes.
\newblock Dimensions of social media addiction among university students in
  kuwait.
\newblock {\em Psychology and Behavioral Sciences}, 4(1):23--28, 2015.

\bibitem{alemneh2021mathematical}
H.~T. Alemneh and N.~Y. Alemu.
\newblock Mathematical modeling with optimal control analysis of social media
  addiction.
\newblock {\em Infectious Disease Modelling}, 6:405--419, 2021.

\bibitem{balakrishnan2013malaysian}
V.~Balakrishnan and A.~Shamim.
\newblock Malaysian facebookers: Motives and addictive behaviours unraveled.
\newblock {\em Computers in Human Behavior}, 29(4):1342--1349, 2013.

\bibitem{bhattacharya2019viral}
S.~Bhattacharya, K.~Gaurav, and S.~Ghosh.
\newblock Viral marketing on social networks: An epidemiological perspective.
\newblock {\em Physica A: Statistical Mechanics and its Applications},
  525:478--490, 2019.

\bibitem{blachnio2019cultural}
A.~B{\l}achnio, A.~Przepi{\'o}rka, O.~Gorbaniuk, M.~Benvenuti, A.~M. Ciobanu,
  E.~Senol-Durak, M.~Durak, M.~N. Giannakos, E.~Mazzoni, I.~O. Pappas, et~al.
\newblock Cultural correlates of internet addiction.
\newblock {\em Cyberpsychology, Behavior, and Social Networking},
  22(4):258--263, 2019.

\bibitem{blachnio2017role}
A.~B{\l}achnio, A.~Przepiorka, E.~Senol-Durak, M.~Durak, and L.~Sherstyuk.
\newblock The role of personality traits in facebook and internet addictions: A
  study on polish, turkish, and ukrainian samples.
\newblock {\em Computers in Human Behavior}, 68:269--275, 2017.

\bibitem{clement2020daily}
J.~Clement.
\newblock Daily time spent on social networking by internet users worldwide
  from 2012 to 2019.
\newblock {\em statista. com [Internet]. Available: https://www-statista-com.
  proxy. bnl. lu/statistics/433871/daily-social-media-usage-worldwide}, 2020.

\bibitem{echeburua2010severe}
E.~Echebur{\'u}a, P.~J. Amor, I.~Loinaz, and P.~de~Corral.
\newblock Severe intimate partner violence risk prediction scale-revised.
\newblock {\em Psicothema}, 22(4):1054--1060, 2010.

\bibitem{elphinston2011time}
R.~A. Elphinston and P.~Noller.
\newblock Time to face it! facebook intrusion and the implications for romantic
  jealousy and relationship satisfaction.
\newblock {\em Cyberpsychology, behavior, and social networking},
  14(11):631--635, 2011.

\bibitem{fang2015global}
B.~Fang, X.-Z. Li, M.~Martcheva, and L.-M. Cai.
\newblock Global asymptotic properties of a heroin epidemic model with
  treat-age.
\newblock {\em Applied Mathematics and Computation}, 263:315--331, 2015.

\bibitem{fox2015dark}
J.~Fox and J.~J. Moreland.
\newblock The dark side of social networking sites: An exploration of the
  relational and psychological stressors associated with facebook use and
  affordances.
\newblock {\em Computers in human behavior}, 45:168--176, 2015.

\bibitem{gaurav2022purchase}
K.~Gaurav, S.~Bhattacharya, Y.~N. Singh, and S.~Ghosh.
\newblock To purchase or to pirate: Investigating the role of social influence
  on digital piracy contagion.
\newblock {\em Pramana}, 96(3):120, 2022.

\bibitem{ghosh2020ensuring}
S.~Ghosh, K.~Gaurav, S.~Bhattacharya, and Y.~N. Singh.
\newblock Ensuring the spread of referral marketing campaigns: a quantitative
  treatment.
\newblock {\em Scientific Reports}, 10(1):11072, 2020.

\bibitem{griffiths2000does}
M.~Griffiths.
\newblock Does internet and computer" addiction" exist? some case study
  evidence.
\newblock {\em CyberPsychology and Behavior}, 3(2):211--218, 2000.

\bibitem{huang2013note}
G.~Huang and A.~Liu.
\newblock A note on global stability for a heroin epidemic model with
  distributed delay.
\newblock {\em Applied Mathematics Letters}, 26(7):687--691, 2013.

\bibitem{j2014internet}
D.~J~Kuss, M.~D~Griffiths, L.~Karila, and J.~Billieux.
\newblock Internet addiction: A systematic review of epidemiological research
  for the last decade.
\newblock {\em Current pharmaceutical design}, 20(25):4026--4052, 2014.

\bibitem{koc2013facebook}
M.~Koc and S.~Gulyagci.
\newblock Facebook addiction among turkish college students: The role of
  psychological health, demographic, and usage characteristics.
\newblock {\em Cyberpsychology, Behavior, and Social Networking},
  16(4):279--284, 2013.

\bibitem{kuss2011online}
D.~J. Kuss and M.~D. Griffiths.
\newblock Online social networking and addiction—a review of the
  psychological literature.
\newblock {\em International journal of environmental research and public
  health}, 8(9):3528--3552, 2011.

\bibitem{lahrouz2011deterministic}
A.~Lahrouz, L.~Omari, D.~Kiouach, and A.~Belmaati.
\newblock Deterministic and stochastic stability of a mathematical model of
  smoking.
\newblock {\em Statistics \& Probability Letters}, 81(8):1276--1284, 2011.

\bibitem{liu2022social}
C.~Liu and J.~Ma.
\newblock Social media addiction and burnout: The mediating roles of envy and
  social media use anxiety.
\newblock In {\em Key Topics in Technology and Behavior}, pages 1--9. Springer,
  2022.

\bibitem{marino2017objective}
C.~Marino, L.~Finos, A.~Vieno, M.~Lenzi, and M.~M. Spada.
\newblock Objective facebook behaviour: Differences between problematic and
  non-problematic users.
\newblock {\em Computers in Human Behavior}, 73:541--546, 2017.

\bibitem{marino2018associations}
C.~Marino, G.~Gini, A.~Vieno, and M.~M. Spada.
\newblock The associations between problematic facebook use, psychological
  distress and well-being among adolescents and young adults: A systematic
  review and meta-analysis.
\newblock {\em Journal of Affective Disorders}, 226:274--281, 2018.

\bibitem{meral2021social}
K.~Z. Meral.
\newblock Social media short video-sharing tiktok application and ethics: data
  privacy and addiction issues.
\newblock In {\em Multidisciplinary approaches to ethics in the digital era},
  pages 147--165. IGI Global, 2021.

\bibitem{omar2020influence}
S.~Z. Omar, Z.~Zaremohzzabieh, A.~A. Samah, J.~Bolong, and H.~A.~M. Shaffril.
\newblock Influence of different facets of internet addiction on subjective
  well-being in malaysia: a comparison across ethnic groups.
\newblock {\em J. Komun. Malays. J. Commun}, 36(2), 2020.

\bibitem{pantic2014online}
I.~Pantic.
\newblock Online social networking and mental health.
\newblock {\em Cyberpsychology, Behavior, and Social Networking},
  17(10):652--657, 2014.

\bibitem{pantic2012association}
I.~Pantic, A.~Damjanovic, J.~Todorovic, D.~Topalovic, D.~Bojovic-Jovic,
  S.~Ristic, and S.~Pantic.
\newblock Association between online social networking and depression in high
  school students: behavioral physiology viewpoint.
\newblock {\em Psychiatria Danubina}, 24(1.):90--93, 2012.

\bibitem{przepiorka2016time}
A.~Przepiorka and A.~Blachnio.
\newblock Time perspective in internet and facebook addiction.
\newblock {\em Computers in Human Behavior}, 60:13--18, 2016.

\bibitem{sanchez2007drinking}
F.~S{\'a}nchez, X.~Wang, C.~Castillo-Ch{\'a}vez, D.~M. Gorman, and P.~J.
  Gruenewald.
\newblock Drinking as an epidemic—a simple mathematical model with recovery
  and relapse.
\newblock In {\em Therapist's Guide to Evidence-Based Relapse Prevention},
  pages 353--368. Elsevier, 2007.

\bibitem{saputri2023social}
R.~A.~M. Saputri and T.~Yumarni.
\newblock Social media addiction and mental health among university students
  during the covid-19 pandemic in indonesia.
\newblock {\em International Journal of Mental Health and Addiction},
  21(1):96--110, 2023.

\bibitem{saqib2022social}
N.~Saqib and F.~Amin.
\newblock Social media addiction: A review on scale development.
\newblock {\em Management and Labour Studies}, 47(3):350--360, 2022.

\bibitem{smith2018social}
A.~Smith and M.~Anderson.
\newblock Social media use in 2018.
\newblock {\em Advances in Applied Sociology}, 10(11), 2018.

\bibitem{soh2018parents}
P.~C.-H. Soh, K.~W. Chew, K.~Y. Koay, and P.~H. Ang.
\newblock Parents vs peers’ influence on teenagers’ internet addiction and
  risky online activities.
\newblock {\em Telematics and Informatics}, 35(1):225--236, 2018.

\bibitem{spertus2005evaluating}
E.~Spertus, M.~Sahami, and O.~Buyukkokten.
\newblock Evaluating similarity measures: a large-scale study in the orkut
  social network.
\newblock In {\em Proceedings of the eleventh ACM SIGKDD international
  conference on Knowledge discovery in data mining}, pages 678--684, 2005.

\bibitem{starcevic2013internet}
V.~Starcevic.
\newblock Is internet addiction a useful concept?
\newblock {\em Australian \& New Zealand Journal of Psychiatry}, 47(1):16--19,
  2013.

\bibitem{turel2019social}
O.~Turel and I.~Vaghefi.
\newblock Social media detox: Relapse predictors.
\newblock {\em Psychiatry research}, 112488, 2019.

\bibitem{verduyn2021impact}
P.~Verduyn, N.~Gugushvili, and E.~Kross.
\newblock The impact of social network sites on mental health: distinguishing
  active from passive use.
\newblock {\em World Psychiatry}, 20(1):133, 2021.

\bibitem{verduyn2020social}
P.~Verduyn, N.~Gugushvili, K.~Massar, K.~T{\"a}ht, and E.~Kross.
\newblock Social comparison on social networking sites.
\newblock {\em Current opinion in psychology}, 36:32--37, 2020.

\bibitem{wang2011epidemiological}
L.~Wang and B.~C. Wood.
\newblock An epidemiological approach to model the viral propagation of memes.
\newblock {\em Applied Mathematical Modelling}, 35(11):5442--5447, 2011.

\bibitem{wegmann2016internet}
E.~Wegmann and M.~Brand.
\newblock Internet-communication disorder: It's a matter of social aspects,
  coping, and internet-use expectancies.
\newblock {\em Frontiers in psychology}, 7:1747, 2016.

\bibitem{wolniczak2013association}
I.~Wolniczak, J.~A. C{\'a}ceres-DelAguila, G.~Palma-Ardiles, K.~J. Arroyo,
  R.~Sol{\'\i}s-Visscher, S.~Paredes-Yauri, K.~Mego-Aquije, and
  A.~Bernabe-Ortiz.
\newblock Association between facebook dependence and poor sleep quality: a
  study in a sample of undergraduate students in peru.
\newblock {\em PloS one}, 8(3):e59087, 2013.

\bibitem{xanidis2016association}
N.~Xanidis and C.~M. Brignell.
\newblock The association between the use of social network sites, sleep
  quality and cognitive function during the day.
\newblock {\em Computers in human behavior}, 55:121--126, 2016.

\bibitem{zhao2012sihr}
L.~Zhao, J.~Wang, Y.~Chen, Q.~Wang, J.~Cheng, and H.~Cui.
\newblock Sihr rumor spreading model in social networks.
\newblock {\em Physica A: Statistical Mechanics and its Applications},
  391(7):2444--2453, 2012.

\bibitem{zhu2022pattern}
L.~Zhu and L.~He.
\newblock Pattern formation in a reaction--diffusion rumor propagation system
  with allee effect and time delay.
\newblock {\em Nonlinear Dynamics}, pages 1--23, 2022.

\end{thebibliography}


\begin{thebibliography}{10}
\providecommand{\url}[1]{{#1}}
\providecommand{\urlprefix}{URL }
\expandafter\ifx\csname urlstyle\endcsname\relax
  \providecommand{\doi}[1]{DOI~\discretionary{}{}{}#1}\else
  \providecommand{\doi}{DOI~\discretionary{}{}{}\begingroup
  \urlstyle{rm}\Url}\fi

\bibitem{alhazzani2020surviving}
Alhazzani, W., M{\o}ller, M.H., Arabi, Y.M., Loeb, M., Gong, M.N., Fan, E.,
  Oczkowski, S., Levy, M.M., Derde, L., Dzierba, A., et~al.: Surviving {S}epsis
  {C}ampaign: guidelines on the management of critically ill adults with
  {C}oronavirus disease 2019 ({COVID}-19).
\newblock Intensive care medicine pp. 1--34 (2020)

\bibitem{althouse2012synchrony}
Althouse, B.M., Lessler, J., Sall, A.A., Diallo, M., Hanley, K.A., Watts, D.M.,
  Weaver, S.C., Cummings, D.A.: Synchrony of sylvatic dengue isolations: a
  multi-host, multi-vector sir model of dengue virus transmission in senegal.
\newblock PLoS Negl Trop Dis \textbf{6}(11), e1928 (2012)

\bibitem{anderson1992infectious}
Anderson, R.M., May, R.M.: Infectious diseases of humans: dynamics and control.
\newblock Oxford university press (1992)

\bibitem{bai2020presumed}
Bai, Y., Yao, L., Wei, T., Tian, F., Jin, D.Y., Chen, L., Wang, M.: Presumed
  asymptomatic carrier transmission of {COVID}-19.
\newblock Jama \textbf{323}(14), 1406--1407 (2020)

\bibitem{bauch2005dynamically}
Bauch, C.T., Lloyd-Smith, J.O., Coffee, M.P., Galvani, A.P.: Dynamically
  modeling sars and other newly emerging respiratory illnesses: past, present,
  and future.
\newblock Epidemiology pp. 791--801 (2005)

\bibitem{behncke2000optimal}
Behncke, H.: Optimal control of deterministic epidemics.
\newblock Optimal control applications and methods \textbf{21}(6), 269--285
  (2000)

\bibitem{bhattacharya2019viral}
Bhattacharya, S., Gaurav, K., Ghosh, S.: Viral marketing on social networks: An
  epidemiological perspective.
\newblock Physica A: Statistical Mechanics and its Applications \textbf{525},
  478--490 (2019)

\bibitem{chatterjee2020studying}
Chatterjee, S., Sarkar, A., Chatterjee, S., Karmakar, M., Paul, R.: Studying
  the progress of covid-19 outbreak in india using sird model.
\newblock medRxiv  (2020)

\bibitem{cheng2020kidney}
Cheng, Y., Luo, R., Wang, K., Zhang, M., Wang, Z., Dong, L., Li, J., Yao, Y.,
  Ge, S., Xu, G.: Kidney disease is associated with in-hospital death of
  patients with {COVID}-19.
\newblock Kidney International  (2020)

\bibitem{chopard1998cellular}
Chopard, B., Droz, M.: Cellular automata, vol.~1.
\newblock Springer (1998)

\bibitem{cucinotta2020declares}
Cucinotta, D., Vanelli, M.: {WHO} declares {COVID}-19 a pandemic.
\newblock Acta bio-medica: Atenei Parmensis \textbf{91}(1), 157--160 (2020)

\bibitem{davies1995effect}
Davies, C.: The effect of neighbourhood on the kinetics of a cellular automaton
  recrystallisation model.
\newblock Scripta metallurgica et materialia \textbf{33}(7), 1139--1143 (1995)

\bibitem{dean2014discrete}
Dean, D.O., Bauer, D.J., Shanahan, M.J.: A discrete-time {M}ultiple {E}vent
  {P}rocess {S}urvival mixture ({MEPSUM}) model.
\newblock Psychological methods \textbf{19}(2), 251 (2014)

\bibitem{diekmann2000mathematical}
Diekmann, O., Heesterbeek, J.A.P.: Mathematical epidemiology of infectious
  diseases: model building, analysis and interpretation, vol.~5.
\newblock John Wiley \& Sons (2000)

\bibitem{van2020aerosol}
van Doremalen, N., Bushmaker, T., Morris, D.H., Holbrook, M.G., Gamble, A.,
  Williamson, B.N., Tamin, A., Harcourt, J.L., Thornburg, N.J., Gerber, S.I.,
  et~al.: Aerosol and surface stability of {SARS}-{CoV}-2 as compared with
  {SARS}-{CoV}-1.
\newblock New England Journal of Medicine \textbf{382}(16), 1564--1567 (2020)

\bibitem{ghosh2020data}
Ghosh, S., Bhattacharya, S.: A data-driven understanding of covid-19 dynamics
  using sequential genetic algorithm based probabilistic cellular automata.
\newblock Applied Soft Computing \textbf{96}, 106692 (2020)

\bibitem{han2020digestive}
Han, C., Duan, C., Zhang, S., Spiegel, B., Shi, H., Wang, W., Zhang, L., Lin,
  R., Liu, J., Ding, Z., et~al.: Digestive symptoms in {COVID}-19 patients with
  mild disease severity: clinical presentation, stool viral {RNA} testing, and
  outcomes.
\newblock The American journal of gastroenterology  (2020)

\bibitem{he2020seir}
He, S., Peng, Y., Sun, K.: Seir modeling of the covid-19 and its dynamics.
\newblock Nonlinear Dynamics pp. 1--14 (2020)

\bibitem{hethcote1973asymptotic}
Hethcote, H.W.: Asymptotic behavior in a deterministic epidemic model.
\newblock Bulletin of Mathematical Biology \textbf{35}, 607--614 (1973)

\bibitem{jin2020epidemiological}
Jin, X., Lian, J.S., Hu, J.H., Gao, J., Zheng, L., Zhang, Y.M., Hao, S.R., Jia,
  H.Y., Cai, H., Zhang, X.L., et~al.: Epidemiological, clinical and virological
  characteristics of 74 cases of coronavirus-infected disease 2019 ({COVID}-19)
  with gastrointestinal symptoms.
\newblock Gut \textbf{69}(6), 1002--1009 (2020)

\bibitem{kermack1927contribution}
Kermack, W.O., McKendrick, A.G.: A contribution to the mathematical theory of
  epidemics.
\newblock Proceedings of the royal society of london. Series A, Containing
  papers of a mathematical and physical character \textbf{115}(772), 700--721
  (1927)

\bibitem{kumar2020covid}
Kumar, A., Nayar, K.R.: {COVID} 19 and its mental health consequences.
\newblock J Ment Heal \textbf{8237}, 1--2 (2020)

\bibitem{lauer2020incubation}
Lauer, S.A., Grantz, K.H., Bi, Q., Jones, F.K., Zheng, Q., Meredith, H.R.,
  Azman, A.S., Reich, N.G., Lessler, J.: The incubation period of coronavirus
  disease 2019 (covid-19) from publicly reported confirmed cases: estimation
  and application.
\newblock Annals of internal medicine \textbf{172}(9), 577--582 (2020)

\bibitem{liu2020reproductive}
Liu, Y., Gayle, A.A., Wilder-Smith, A., Rockl{\"o}v, J.: The reproductive
  number of {COVID}-19 is higher compared to {SARS} coronavirus.
\newblock Journal of travel medicine  (2020)

\bibitem{mairesse2014around}
Mairesse, J., Marcovici, I.: Around probabilistic cellular automata.
\newblock Theoretical Computer Science \textbf{559}, 42--72 (2014)

\bibitem{murthy2020care}
Murthy, S., Gomersall, C.D., Fowler, R.A.: Care for critically ill patients
  with {COVID}-19.
\newblock Jama \textbf{323}(15), 1499--1500 (2020)

\bibitem{nacoti2020epicenter}
Nacoti, M., Ciocca, A., Giupponi, A., Brambillasca, P., Lussana, F., Pisano,
  M., Goisis, G., Bonacina, D., Fazzi, F., Naspro, R., et~al.: At the epicenter
  of the {COVID}-19 pandemic and humanitarian crises in {I}taly: changing
  perspectives on preparation and mitigation.
\newblock NEJM Catalyst Innovations in Care Delivery \textbf{1}(2) (2020)

\bibitem{nishiura2020estimation}
Nishiura, H., Kobayashi, T., Miyama, T., Suzuki, A., Jung, S., Hayashi, K.,
  Kinoshita, R., Yang, Y., Yuan, B., Akhmetzhanov, A.R., et~al.: Estimation of
  the asymptomatic ratio of novel coronavirus infections ({COVID}-19).
\newblock medRxiv  (2020)

\bibitem{pan2020clinical}
Pan, L., Mu, M., Yang, P., Sun, Y., Wang, R., Yan, J., Li, P., Hu, B., Wang,
  J., Hu, C., et~al.: Clinical characteristics of {COVID}-19 patients with
  digestive symptoms in {H}ubei, {C}hina: a descriptive, cross-sectional,
  multicenter study.
\newblock The American journal of gastroenterology \textbf{115} (2020)

\bibitem{rocklov2020covid}
Rockl{\"o}v, J., Sj{\"o}din, H., Wilder-Smith, A.: {COVID}-19 outbreak on the
  {D}iamond {P}rincess cruise ship: estimating the epidemic potential and
  effectiveness of public health countermeasures.
\newblock Journal of travel medicine \textbf{27}(3), taaa030 (2020)

\bibitem{sacks1977transition}
Sacks, S.T., Chiang, C.L.: A transition-probability model for the study of
  chronic diseases.
\newblock Mathematical Biosciences \textbf{34}(3-4), 325--346 (1977)

\bibitem{sante2010cellular}
Sant{\'e}, I., Garc{\'\i}a, A.M., Miranda, D., Crecente, R.: Cellular automata
  models for the simulation of real-world urban processes: {A} review and
  analysis.
\newblock Landscape and Urban Planning \textbf{96}(2), 108--122 (2010)

\bibitem{shim2020transmission}
Shim, E., Tariq, A., Choi, W., Lee, Y., Chowell, G.: Transmission potential and
  severity of {COVID}-19 in {S}outh {K}orea.
\newblock International Journal of Infectious Diseases  (2020)

\bibitem{spinelli2020covid}
Spinelli, A., Pellino, G.: {COVID}-19 pandemic: perspectives on an unfolding
  crisis.
\newblock The British Journal of Surgery  (2020)

\bibitem{toffoli1987cellular}
Toffoli, T., Margolus, N.: Cellular automata machines: a new environment for
  modeling.
\newblock MIT press (1987)

\bibitem{torales2020outbreak}
Torales, J., O’Higgins, M., Castaldelli-Maia, J.M., Ventriglio, A.: The
  outbreak of {COVID}-19 coronavirus and its impact on global mental health.
\newblock International Journal of Social Psychiatry p. 0020764020915212 (2020)

\bibitem{wang2020immediate}
Wang, C., Pan, R., Wan, X., Tan, Y., Xu, L., Ho, C.S., Ho, R.C.: Immediate
  psychological responses and associated factors during the initial stage of
  the 2019 coronavirus disease ({COVID}-19) epidemic among the general
  population in {C}hina.
\newblock International journal of environmental research and public health
  \textbf{17}(5), 1729 (2020)

\bibitem{wolfram2018cellular}
Wolfram, S.: Cellular automata and complexity: collected papers.
\newblock CRC Press (2018)

\bibitem{yu2020familial}
Yu, P., Zhu, J., Zhang, Z., Han, Y.: A familial cluster of infection associated
  with the 2019 novel coronavirus indicating possible person-to-person
  transmission during the incubation period.
\newblock The Journal of infectious diseases \textbf{221}(11), 1757--1761
  (2020)

\bibitem{zheng2020covid}
Zheng, Y.Y., Ma, Y.T., Zhang, J.Y., Xie, X.: {COVID-19} and the cardiovascular
  system.
\newblock Nature Reviews Cardiology \textbf{17}(5), 259--260 (2020)

\end{thebibliography}



\end{document}